# Importance Sampling via Variational Optimization

**Ydo Wexler**
Computer Science Dept.
Technion, Haifa 32000, Israel
ywex@cs.technion.ac.il

**Dan Geiger**
Computer Science Dept.
Technion, Haifa 32000, Israel
dang@cs.technion.ac.il

## Abstract

Computing the exact likelihood of data in large Bayesian networks consisting of thousands of vertices is often a difficult task. When these models contain many deterministic conditional probability tables and when the observed values are extremely unlikely even alternative algorithms such as variational methods and stochastic sampling often perform poorly. We present a new importance sampling algorithm for Bayesian networks which is based on variational techniques. We use the updates of the importance function to predict whether the stochastic sampling converged above or below the true likelihood, and change the proposal distribution accordingly. The validity of the method and its contribution to convergence is demonstrated on hard networks of large genetic linkage analysis tasks.

## 1 Introduction

Stochastic sampling methods in Bayesian networks are approximation algorithms that generate instantiations of the network according to the probabilities of a sampling model [2, 12, 13, 16, 17, 23, 28]. Unlike exact inference which is exponential in the treewidth of the network, sampling execution time is in general independent of the topology of the network and usually linear in the size of the network and the number of samples generated. On the other hand, precision of the sampling estimate tends to increase with the number of samples generated, but decreases with the size of the network. Therefore, while in small networks it may be possible to instantiate almost all the sample space and quickly converge to exact probabilities, in very large networks only a small fraction of the total sample space can be explored within a reasonable time, and achieving a good precision requires more resources.

A problem that remains largely unresolved in sampling is to compute the likelihood of evidence in large networks with many deterministic conditional probabilities and extremely unlikely evidence. Leading sampling algorithms devised to overcome the problem of unlikely evidence [2, 28] sometime suffer from low convergence rates when the network contains deterministic probabilities and may not always yield solutions close to the true likelihood.

In this paper we present a new importance sampling algorithm for Bayesian networks that is specifically designed to work well in networks with many deterministic conditional probabilities. The algorithm first employs known variational techniques [1, 14, 15, 18, 22, 26] in order to approximate the joint probability when using a simplified network in which some of the edges of the original network were removed. This is followed by exact inference on the simplified network to achieve an initial proposal distribution. To refine the algorithm we add an adaptive scheme that allows the algorithm to learn a good sampling function and update it accordingly. This adaptation is performed similar to the simulated annealing framework [19]. Then, we utilize the updates to assess whether the sampling procedure was under or over estimating the true likelihood and amend the proposal distribution accordingly. To demonstrate the power of our algorithm we applied it to large networks with many deterministic probabilities that model genetic linkage analysis problems. We compare the results with other state-of-the-art sampling algorithms [2, 23] along with an MCMC software specially designed for genetic linkage tasks [24], and demonstrate a clear improvement in estimation.

## 2 Background

We start with a brief overview of the importance sampling method and algorithms devised in this frame-work. We follow the generic importance sampling algorithm for Bayesian networks as given in [2]. Then we explain the choice of the Kullback-Liebler (KL) divergence as a measure of the quality of the importance function, and proceed with a short survey of algorithms that dynamically update the importance function.

## 2.1 Importance sampling and the KL divergence

Importance sampling (IS) is a powerful technique for rare event simulation that can be used in the Monte Carlo method. Assume we have a joint distribution $P(X)$, in the form of a Bayesian network, over a set of discrete variables $X$, and we wish to compute the likelihood of evidence $P(E = e) = \sum_h P(H, E = e)$ where $E \subseteq X \setminus H$ are observed variables and $H$ are the unobserved ones. The idea behind the IS method is that certain instances $h \in H$, for which $P(h, e)$ is large, have more impact on the likelihood estimated than other instances. Therefore, IS samples with a biased proposal distribution $Q(H)$, called the importance function, that prefers to sample more "influential" instances. In order to obtain an unbiased estimator, the simulation outputs are weighted inversely to $Q(H)$, and the estimate after $M$ samples is given by

$$\tilde{P}(E = e) = \frac{1}{M} \sum_{i=1}^{M} \frac{P(h_i, e)}{Q(h_i)} \quad (1)$$

where $h_i$ is the instantiation of variables $H$ in the $i^{th}$ sample [13, 16, 23, 25]. First, we note that since $\tilde{P}(E = e)$ is an unbiased estimator its expected value equals $P(e)$. In addition, it is easy to show that as $M$ grows to infinity the estimate converges to the true likelihood and its variance decreases linearly with $M$ [8].

A fundamental issue in implementing importance sampling is the choice of an importance function $Q(H)$ that guarantees variance reduction. The convergence rate of the method critically depends on the variance, and while a good importance function can lead to significant run-time savings, an inflated variance can lead to slow convergence. Being an any-time algorithm, the risk lies in ending the simulation prior to converging sufficiently close to the true likelihood, yielding incorrect solutions.

The optimal importance function for calculating $P(E = e)$ via Eq. 1 is $Q = P(H|e)$, as shown in [21]. This distribution is proportional to $P(H, e)$ and when using it we get zero variance as it yields $\frac{P(h,e)}{P(h|e)} = P(e)$ for every sample $h$. However, one practical requirement of $Q(H)$ is that it should be easy to sample from. Finding the optimal proposal distribution $P(H|e)$ is as hard as the original problem of computing $P(e)$. However, as many studies pointed out, using distributions close to the posterior can still yield a low variance sampling, which in turn leads to precise sampling simulations, and can save computational resources and time.

Several sampling and in particular importance sampling techniques use the Kullback-Liebler divergence to quantify the closeness of $Q(H)$ to the posterior distribution [5, 9, 21]. The Kullback-Liebler divergence between the two distributions $Q(H)$ and $P(H|E = e)$ is given by:

$$D(Q(H) \| P(H|E = e)) = \sum_h Q(h) \log \frac{Q(h)}{P(h|e)}.$$

One reason to use the KL divergence as the measure for accuracy of $Q$ is its close connection to the expected variance of the samples. We now provide a novel formalization for this known connection. Consider the weighted power mean $M_w^r(Z)$ of a series of real numbers $Z = \{z_1, \ldots, z_n\}$ defined for every real $r \in \mathbb{R}$ as

$$M_w^r(z_1, \ldots, z_n) = \begin{cases} \left[\sum_{i=1}^n w_i z_i^r\right]^{1/r} & \text{if } r \neq 0 \\ \prod_{i=1}^n z_i^{w_i} & \text{if } r = 0 \end{cases}$$

where $w_1, \ldots, w_n$ are positive real numbers such that $\sum_{i=1}^n w_i = 1$. The KL divergence, $-D(Q(H) \| P(H, e))$, between $Q$ and the non-normalized function $P(H, E = e)$ is the logarithm of the zeroth $(r = 0)$ power mean of the series $\frac{P(h_1,e)}{Q(h_1)}, \frac{P(h_2,e)}{Q(h_2)}, \ldots$ weighted by $Q(h_1), Q(h_2), \ldots$. On the other hand, the expected variance with respect to $P(e)$ when sampling according to $Q$ equals $\left(M_Q^2(\frac{P}{Q})\right)^2 - \left(M_Q^1(\frac{P}{Q})\right)^2 = \left(M_Q^2(\frac{P}{Q})\right)^2 - (P(e))^2$. We also note that the power mean inequality $M_Q^0 \leq M_Q^2$ holds for every normalized distribution $Q$. Now, for two importance distributions $Q_1$ and $Q_2$, for which the ratios $r_Q = \frac{M_Q^2(\frac{P}{Q})}{M_Q^0(\frac{P}{Q})}$ differ only by a multiplicative constant $c$, a small difference in the KL divergence $0 < d = D(Q_1 \| P) - D(Q_2 \| P)$, that satisfies $d > \ln c$, affects the expected variance exponentially since

$$\frac{E_{Q_1}[Var(\frac{P}{Q})]}{E_{Q_2}[Var(\frac{P}{Q})]} = \frac{\left(M_{Q_1}^2\right)^2 - (P(e))^2}{\left(M_{Q_2}^2\right)^2 - (P(e))^2} \quad (2)$$

$$= \frac{\left(r_{Q_1} M_{Q_1}^0\right)^2 - (P(e))^2}{\left(r_{Q_2} M_{Q_2}^0\right)^2 - (P(e))^2} = \frac{\left(r_{Q_1} M_{Q_2}^0\right)^2 e^{2d} - (P(e))^2}{\left(r_{Q_2} M_{Q_2}^0\right)^2 - (P(e))^2} \geq \frac{e^{2d}}{c^2}$$

This means that a small improvement in the KL divergence can yield a dramatic drop in variance and thus improving the accuracy of $\tilde{P}(e)$.

A second reason to use the KL divergence is that this measure can be easily computed either analytically, for simple importance functions, or statistically for more complex ones, as we now show. By taking an average of $\log \frac{P(h,e)}{Q(h)}$ over the samples $h$, the KL divergence is derived via:

$$\frac{\sum_{i=1}^{M} \log \frac{P(h_i)}{Q(h_i)}}{M} \stackrel{M \to \infty}{\longrightarrow} E_Q[\log \frac{P(e,h)}{Q(h)}] \quad (3)$$

$$= \sum_h Q(h) \log \frac{P(e,h)}{Q(h)} = \log P(e) - D(Q(H) \| P(H|e)).$$

The mean in Eq. 3 converges fast to its expected value, because its variance equals $\left(M_Q^2(\log \frac{P}{Q})\right)^2 - \left(M_Q^1(\log \frac{P}{Q})\right)^2$ which is of the order of $\log\left(E_Q[Var(\frac{P}{Q})]\right)$.

## 2.2 Static importance sampling algorithms

The first importance sampling algorithms in Bayesian networks were logic sampling [16] and likelihood weighting [13, 23], sampling the network in a topological order using the prior distribution as the importance function. Since in networks with unlikely evidence, the prior tend to be significantly different from the posterior, the two algorithms produce large variance and are inefficient. Backward sampling [12] is an importance sampling algorithm which tries to incorporate evidence into the importance function by generating samples in the reverse-topological order. This algorithm sometime suffers from a mismatch between the importance function and the optimal proposal distribution, and occasionally leads to poor convergence. Hernadez, Moral & Salmeron [17] suggested an algorithm that incorporates evidence into the importance function, by using an approximate inference that resembles the bucket elimination algorithm [7]. In this approximation algorithm, tables in a bucket are not combined if they are to exceed a limiting size. For complex networks, where many table combinations remain unaccounted, much information is lost and the importance function deteriorates.

The evidence pre-propagation importance sampling algorithm (EPIS-BN) [28] employs loopy belief propagation to compute the approximate posterior belief $P(x_i|e)$ for every node $X_i$ in the network. The main idea behind the algorithm is to replace the optimal proposal distribution $P(H|e) = \frac{P(H,e)}{P(e)}$ and sample instead using the importance function

$$Q(H) = \prod_i P(X_i|\text{pa}(X_i), E = e) \quad (4)$$

where $\text{pa}(X_i)$ is the set of parents variables of $X_i$ in the network. It uses the marginal distribution to approximate the desired function $Q$ and applies a heuristic that increases probabilities of very small value. According to the reported results, this algorithm exhibits the best convergence properties on networks with extremely unlikely evidence. However, as the authors point out, when the posterior probabilities change dramatically as a result of the evidence the suggested importance function may perform poorly. This often happens when the network contains many deterministic conditional probabilities. For example, in the simple network in Fig. 1 where the evidence is $D = 0$, the instance $h = \{A = 0, B = 0, C = 0\}$ has probability zero. Still, using the probabilities $P(x|e)$ to sample the network will result in sampling the instance $h$ approximately 6.5% of the times. Other importance sampling algorithms update the importance function during simulation, and we discuss this approach in the following section.

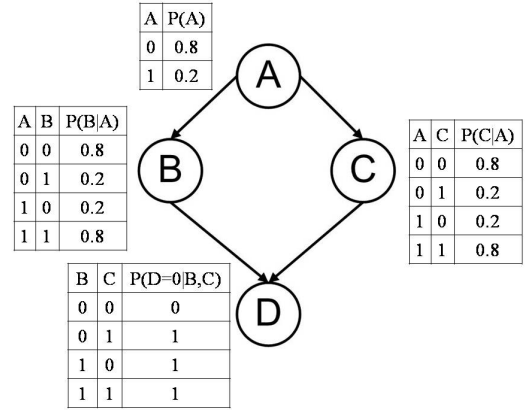

Figure 1: A small Bayesian network in which the evidence is $D = 0$. Here the probability $P(a, b, c|d)$ is utterly different from $P(a|d)P(b|a,d)P(c|a,d)$ for some instantiations like (A=0,B=0,C=0).

## 2.3 Adaptive importance sampling algorithms

The difficulty in setting a close to optimal importance function $Q$ led to methods that try to estimate the properties of the optimal function from the samples already generated. These estimations are used to update the importance function and to close the gap from the optimal function. In addition, some of these adaptive methods, supported by the assumption that the simulation becomes more accurate with each update of the importance function, also monotonically increase the weight associated with each sample. Denoting by $Q_k$ the importance function after $k$ updates, by $w_k$ the weight associated with samples generated in the $k^{th}$ iteration, and by $M_k$ the number of samples in the $k^{th}$ iteration, the estimated likelihood can be written as follows:

$$\tilde{P}(E = e) = \frac{\sum_k \sum_{i=1}^{M_k} w_k \frac{P(h_i, y)}{Q_k(h_i)}}{\sum_k M_k w_k} \quad (5)$$

We note that in most cases the adaptive behavior bias the estimator, namely, the methods can no longer guarantee that the expected estimate equals $P(e)$.

The algorithm self-importance sampling (SIS) [23] instantiates each node of the network given the states of the parent nodes in a topological order, and in each iteration updates the conditional probability tables according to the samples generated in that iteration. The purpose of this update is to ensure that the proposal distribution gets gradually closer to the posterior distribution, although no exact measure for this closeness is given.

The Cross-Entropy (CE) method [4] is another adaptive importance sampling procedure, which was devised for general optimization and rare-event simulation. In the $k^{th}$ iteration the method uses only an "elite" part $S_k$ of the samples generated at that iteration to update the importance function $Q_k$ via the optimization equation:

$$Q_k = \text{argmax}_Q \sum_{h \in S_k} P(h, e) \log Q(h)$$

This elite part is the top $\rho$ portion of the samples generated in terms of $P(h, e)$, for a user defined $\rho \in [0, 1]$. It is guaranteed that for large enough number of iterations the method will converge to the true probability, however, this convergence may be slow.

An additional algorithm, adaptive importance sampling (AIS-BN) [2], was devised to sample Bayesian networks with extreme unlikely evidence and is considered among the fastest converging importance sampling algorithms. The main idea behind the algorithm is similar to the one behind EPIS-BN, namely to replace the optimal proposal distribution with the function given by Eq. 4. The difference is that AIS-BN updates the importance function according to rules based on gradient descent, and in addition, initializes the probability of parents of evidence nodes to the uniform distribution. We note that in Bayesian networks with many deterministic conditional probabilities and evidence almost exclusively in the leaves, such as those that model genetic linkage problems, the initialization rules induce almost a uniform importance function, far from the posterior distribution. In addition, on such networks the algorithm sometimes converges to importance functions far from the optimum.

## 3 Variational Sampling

We now consider the problem of setting a useful proposal distribution $Q(H)$ in probabilistic models. First, we suggest an importance sampling algorithm that employs variational techniques [15, 22] and exact inference procedures on a simplified network. Then, we refine the sampling algorithm and provide an adaptation scheme that aims to reduce the KL divergence through annealing-like adaptation. Finally, we observe that intermediate lower bounds on the likelihood are correlated with the sampling estimates, and use this observation to further improve importance sampling.

### 3.1 Setting importance distributions

The use of simplified models that are more amenable to inference than the original model is the key idea behind variational techniques. In general, these methods aim to lower bound the likelihood of evidence $P(E = e)$ by using Jensen's inequality:

$$\log P(e) \geq \sum_h Q(h) \log \frac{P(h, e)}{Q(h)} \quad (6)$$

where $Q(H)$ is a probability distribution selected by the variational algorithm. To set a high lower bound the distribution $Q$ has to be close to the posterior $P(H|e)$. Hence, the variational approach offers iterative algorithms for finding distributions $Q$ that minimize the KL divergence between $Q$ and the target distribution $P(H|e)$ and in most cases arrive at a stationary point of the KL divergence [1, 3, 14, 26].

A straight forward usage of this approach is to employ the distribution $Q$ suggested by the variational algorithms as an importance function. This can be justified in light of the discussion in Section 2.1, and in particular Eq. 2, claiming that distributions $Q$ close to the target distribution $P$ are expected to yield a low variance sampling. This idea is analogous to the usage of belief propagation in [28] and that of variational methods in [6]. However, for some class of networks, especially complex models with many deterministic probability tables and extremely unlikely evidence, such as those used to model genetic analysis problems [11, 10], these techniques are not sufficiently accurate. They either yield distributions far from the desired posterior distribution [14], or when the form of $Q$ is less constrained, the time required by these procedures is not feasible.

We propose a four phases algorithm, called VarIS, that given a network $N$ sets an importance function $Q$ and an order of sampling. In the first phase edges are deleted yielding a simplified network $N'$ with a reduced treewidth. The network $N'$ factors according to $P'(x) = \prod_i P'(X_i | \text{pa}'(X_i))$ where $\text{pa}'(X_i)$ is the set of parents variables of $X_i$ in $N'$. Then, in the second phase we employ a variational technique that gets as input the networks $N$ and $N'$ after removing evidence nodes from them. The variational procedure outputs a distribution $P'(H)$ that factors according to the topology of $N'$ and approximates the distribution $P(H)$. Note that in this phase the variational procedure reduces Eq. 6 to maximizing $\sum_h P'(h) \log \frac{P(h)}{P'(h)}$ closer to zero. In the third phase we perform an exact inference on the simplified network $N'$ via bucket elimination [7], using an elimination order of convenience. We let the importance function $Q$ factor according to:

$$Q(h) = \prod_i Q_i(x_i|s_i)$$

where $X_i \in X$ and the subset of variables $S_i \subseteq X$ contains all the variables in the $i^{th}$ bucket after eliminating $X_i$. Suppose that bucket $i$ contains the table $\lambda(X_i, S_i)$ before eliminating $X_i$ and the table $\lambda(S_i)$ after this elimination, then for every instance $(x_i, s_i)$ we set $Q_i(x_i|s_i) = \frac{\lambda(x_i, s_i)}{\lambda(s_i)}$. The sampling order is set to the reverse elimination order. The fourth phase reinstates edges that were deleted in the first phase. If the edge $U \to V$ was deleted from $N$ and variable $U$ appears before $V$ in the sampling order, the function $Q_V$ according to which we sample the variable $V$ is multiplied by $\sum_{\text{pa}'(v)} P(v|\text{pa}(v))$ to take into account the dependence between $U$ and $V$ that is lost in $N'$ without increasing the computational burden of the algorithm.

Algorithm VarIS is summarized in Alg. 1 using three external procedures: *DelEdges(N)* which given a Bayesian network $N$ outputs a simplified network $N'$, *VarTech(N, N')*, a standard variational technique, that sets the prior distribution $P'(H)$ of $N'$ close to $P(H)$, and *FindRevOrder(N)* which returns an elimination order of network $N$ in reversed order.

The algorithm's efficiency in setting a good importance function depends on the form of the simplified network $N'$. Deleting the edges in phase 2 is done via the method of Choi & Darwiche [3], but also any other edge deletion criterion such as that suggested by Xing, Jordan & Russell [27] is acceptable. The third phase resembles the method of Hernández et al. [17] in using inference to incorporate the evidence into the sampling function. The difference is that we perform exact inference on a simplified network $N'$ while Hernández et al. [17] perform approximate bucket elimination on the original network $N$.

### 3.2 Annealing adaptation scheme

The sampling algorithm presented so far uses a static importance function. We now present an annealing like adaptive procedure to update the importance sampling function by learning from the samples generated during the simulation in order to further reduce the sampling variance. This approach is similar in nature to the one used by SIS [23] and AIS-BN [2] with the difference that here an estimation of the KL divergence directs the update procedure.

---

**Algorithm 1**: VarIS

**Input**: Bayesian network $N$ with $n$ nodes
     $X = H \cup E$ and tables $\lambda_1, \ldots, \lambda_n$
**Output**: Distribution $Q(H) = \prod_{i=1}^n Q_i(H_i)$ and a sampling order $R$

$R, D$ - Arrays of integers;
$B$ - An array of buckets;
$Q$ - An array of multidimensional tables;
**First phase:** $N' \leftarrow$ *DelEdges(N)*;
**Second phase:** $P'(H) \leftarrow$ *VarTech(N\E, N'\E)*;
**Third phase:**
$R \leftarrow$ *FindRevOrder(N)*;
**for** $i$ from 1 to $n$ **do**
    $D_{R_i} \leftarrow i$
**end**
**Initialization:** associate every table
$\lambda_i = P'(X_i|\text{pa}'(X_i))$ with the bucket $B_j$ where
$j = \max_{V \in \lambda_i} D_V$
**Elimination: for** $k \leftarrow n$ to 1 **do**
    $i \leftarrow R_k$;
    Let $S_i$ be the set of variables aside of $X_i$ in tables
    $\lambda_1, \ldots, \lambda_m$ in $B_i$;
    **for** $S_i = s_i$ **do**
        $\lambda_i(x_i, s_i) = \prod_{j=1}^m \lambda_j$;
    **end**
    **if** (bucket with observed variable) $X_i \in E$ **then**
        set $\lambda_i(x_i, s_i) = 0$ for instances where $x_i$ is
        different from the observed value;
    **end**
    $\lambda_i(s_i) = \sum_{x_i} \lambda_i(x_i, s_i)$;
    $Q_i(x_i|s_i) = \frac{\lambda_i(x_i, s_i)}{\lambda_i(s_i)}$;
    Add $\lambda_i(S_i)$ to the bucket $B_j$ where
    $j = \max_{V \in \lambda_i} D_V$
**end**
**Fourth phase:**
Let $E_{del}$ be the set of deleted edges;
**for** $(U, V) \in E_{del}$ **do**
    **if** ($U$ is sampled before $V$) $D_U < D_V$ **then**
        **for** $(V, U, S_V) = (v, u, s_V)$ **do**
            $Q_V(v|s_V, u) = Q_V(v|s_V) \cdot \sum_{\text{pa}'(v)} P(v|\text{pa}(v))$
        **end**
    **end**
**end**
return $Q, R$

Dividing the simulation into batches, each of $m$ samples, we update the probabilities in $Q$ based on the samples in the recent batch. Denote by $D_k$ the KL divergence in the $k^{th}$ batch. An update occurs if it yields an improvement in the KL divergence, namely that $D_k - D_{k-1} < 0$, or otherwise with probability

$$p = e^{-k\delta_k}$$

where $k$ is the batch number and $\delta_k = D_k - D_{k-1}$. Let the variable $X_i$ be sampled in $Q$ conditioned on the subset of variables $S_i$ that appear earlier in the sampling order, and let $N_k(u)$ be the number of samples in the $k^{th}$ batch where $U = u$, for a set of variables $U$. Then, the update is

$$Q'(x_i|s_i) = (1 - \eta(k))Q(x_i|s_i) + \eta(k)\frac{N_k(x_i, s_i)}{N_k(s_i)}$$

where the mixing rate $\eta(k)$ changes with time, and equals $\eta(k) = \eta_0 \left(\frac{\eta_f}{\eta_0}\right)^{k/k_{max}}$ similar to [2], and where $\eta_0$ is the initial mixing rate and $\eta_f$ is the mixing rate at batch $k_{max}$. The values $Q'(x_i|s_i)$ are normalized after the update.

In addition, we associate weights to the samples according to Eq. 5. The weights are initialized close to zero, and increased reversely proportional to the standard deviation of samples in the iteration.

A particular problem in this adaptive scheme is that sometimes the KL divergence $D(Q \| P)$ can not be analytically computed within a reasonable time, and even if $Q$ is of a simple form it can still prolong the simulation to a large extent. We use Eq. 3 to obtain an approximation of the KL divergence without adding computations via the average of $\log \frac{P(h,e)}{Q(h)}$ over all $m$ samples in the batch. In all runs we performed, this average converged after a few hundreds samples to a value within 0.5% of the true KL value, and the results show that relying on these estimates yields an adaptive scheme that benefits the sampling algorithm.

### 3.3 Directing the importance function

The correlation of the KL divergence and the variance of the sampling estimates $\tilde{P}(e)$ allows the use of a simple method to assess whether the sampling estimate converges above or below the true likelihood and to change the sampling function accordingly.

The idea is that if an importance sampling algorithm for Bayesian networks converges to a value that is below the true likelihood then often it samples mostly instances $h$ for which $\frac{P(h,e)}{Q(h)} < P(e)$. The reason for this phenomenon is that the importance function is sometimes too uniform, namely, that for most pairs of instances $h_1$ and $h_2$ for which $P(h_1, e) - P(h_2, e) = \delta > 0$, the difference $Q(h_1) - Q(h_2) < \delta$. Consequently, most of the time the simulation samples instances with probability $Q(h)$ higher than the optimal, yielding $\frac{P(h,e)}{Q(h)} < P(e)$.

A significant positive correlation between $D_k$ and $\tilde{P}_k(e)$, which is the estimate on $P(e)$ when considering only samples generated in the $k^{th}$ batch, indicates that the simulation converges to a value below the true likelihood. The reasoning is that at the limit $Q$ reaches the optimal function and $\tilde{P}_k(e)$ reaches $P(e)$. Then, from the correlation one can conclude that the values $\tilde{P}_k(e)$ until reaching the limit are smaller than $P(e)$.

The correlation between the last $l$ values $D_k$ and the corresponding values $\tilde{P}_k(e)$ is computed after each batch using Pearson correlation. The KL divergence is calculated via the average of $\log \frac{P(h,e)}{Q(h)}$ as in Section 3.2. When the correlation is significant we change $Q$ so that high probabilities will get larger, and low probabilities will get lower. In particular, we replace every probability $q < \alpha$ in the simplified network $N'$ by $q^{1+\beta}$ and every probability $q > 1-\alpha$ by $q^{1-\beta}$, where $\alpha < 0.5$ and $\beta < 1$ are parameters, and normalize $Q$.

The same reasoning applies when the algorithm converges to a value above the true likelihood, with the difference that then we expect to find a significant negative correlation between $-D(Q(H) \| P(H, e))$ and $\tilde{P}_k(e)$.

## 4 Experimental results

We demonstrate the power of the methods suggested herein by applying them to Bayesian networks that model difficult genetic linkage analysis problems as described in [11, 10]. The parameters we used in our runs are $m = 1000$, $\eta_0 = 0.12$, $\eta_f = 0.03$, $\alpha = 0.1$, $\beta = 0.2$, $l = 10$, and the initial weight $w_0 = 0.001$. These parameters were set after training the algorithm on a few small-size genetic linkage problems that are not included in the results here.

Genetic linkage analysis takes as input a family pedigree in which some individuals are affected with a genetic disease, along with marker readings from affected and/or healthy individuals. The output is the likelihood of data as a function of the location of a disease gene and the given pedigree. Locations yielding maximum or close to maximum likelihood are singled out as suspect regions for further scrutiny. The exact computation of this likelihood is often too complex and approximations are needed.

We applied algorithm VarIS to various Bayesian networks that model real-life genetic linkage prob-lems. We compare the performance of the algorithm with and without the adaptation features with three other sampling algorithms: self importance sampling (SIS) [23], AIS-BN [2], and the software SimWalk2 V2.86 [24] specially designed for the task of genetic linkage analysis. In order to avoid the problem described in Section 2.1 and illustrated via Fig. 1 that most samples have zero probability, we improved the two algorithms SIS and AIS-BN by letting them sample only from the sample space of feasible instances. This change was achieved without damaging essential properties of these algorithms. We also note that without this change both algorithms did not sample even a single instance with a positive probability. The two modified algorithms are denoted SIS* and AIS-BN* respectively.

Fig. 2 shows the error in ln-likelihood for every location of the disease gene when applying the algorithms to networks that model a complex pedigree with 115 individuals studied by Narkis et al.[20]. We considered 20 different locations for the disease gene, examining the linkage of this gene with one marker location at a time (i.e., two-point analysis). Each run generated 100,000 samples from a network with an average of 800 nodes and with an average of more than 99.99% zeroes in the joint probability distribution represented by the network. The exception was the program SimWalk2 that performed faster and thus executed 1,840,000 sampling steps. The exact analysis took several days using the software SUPERLINK V1.5 [10] designed for exact genetic linkage analysis. On the average VarIS run 90 times faster than exact analysis on these 20 examples, taking approximately 120 seconds rather than 3 hours.

To test the added value of the adaptation features described in Sections 3.2 and 3.3 we compared the full method (VarIS) with two variants: one without the adaptation described in Section 3.3 (VarIS-A), and the other also without the annealing adaptation (VarIS-B). The results on the same pedigree used for Fig. 2 are illustrated in Fig. 3. The line %VarIS in the figure means the error in percentage from the exact ln-likelihood value. For example, at location 4 the error in ln-likelihood is 1.73 and the true result is $-110.1$ yielding 1.57% error.

Finally, we applied algorithm VarIS to networks that model the disease gene at a specific location with an increasing number of nearby markers using the data from [20]. We report the standard LOD score for each set of markers, which is the log ratio between the likelihood of the data assuming the disease gene is nearby a marker and the likelihood of the data assuming the disease gene resides on another chromosome. The run time for computing the exact LOD score with one marker was approximately 3 hours with SUPER-

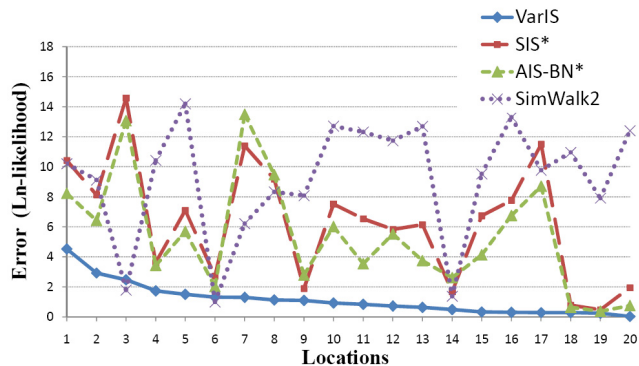

Figure 2: The error in ln-likelihood of 4 sampling algorithms when applied to networks that model genetic analysis of a pedigree studied in [20]. Locations are ordered in decreasing order of errors by VarIS.

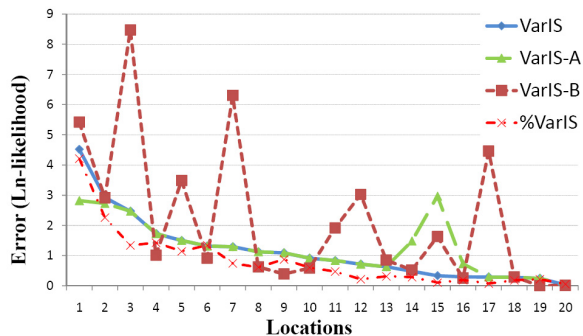

Figure 3: The error in ln-likelihood of 3 variants of VarIS when applied to networks that model genetic analysis of a pedigree studied in [20]. Locations are ordered in decreasing order of errors by VarIS. The line %VarIS represents error in percentage relative to the exact ln-likelihood values.

LINK V1.5, but computing exact LOD scores with two markers (i.e., three-point analysis) or more is not feasible for this pedigree. In Table 1, we detail the computed LOD scores and the average execution times for 1,000,000 samples. The LOD score converges to a high value around 11.3, supporting the hypothesis that this is the location of the disease gene, as proposed in [20]. In addition, the time and space requirements

| # Markers used | 1 | 2 | 3 | 4 | 5 | 6 |
|---|---|---|---|---|---|---|
| LOD | 8.96 | 10.84 | 11.49 | 11.52 | 11.21 | 11.35 |
| time | 1250 | 2719 | 4102 | 5185 | 6956 | 8205 |

Table 1: LOD scores and execution times (in seconds) as a function of the number of markers examined with the disease gene.

for this analysis grow roughly linearly with the number of markers and network size.

## Acknowledgments

We thank Reuven Rubinstien for a helpful discussion. The research is supported by the Israel Science Foundation and the Israeli Science Ministry.